\documentclass[12pt]{article}
\usepackage{epsfig}
\begin{document}
\title{Absorption of $\psi(2S)$ mesons in nuclei}
\author{E. Ya. Paryev\\
{\it Institute for Nuclear Research of the Russian Academy of Sciences}\\
{\it Moscow, Russia}}

\renewcommand{\today}{}
\maketitle

\begin{abstract}
In the present work we explore the inclusive $\psi(2S)$ meson photoproduction  from nuclei near the kinematic threshold within the collision model, based on the nuclear spectral function, for incoherent direct photon--nucleon charmonium creation processes. The model takes into account the final $\psi(2S)$ absorption, target nucleon binding and Fermi motion. We calculate the absolute and relative excitation functions for production of $\psi(2S)$ mesons on $^{12}$C and $^{184}$W target nuclei at near-threshold photon beam energies of 8.0--16.4 GeV, the absolute momentum differential cross sections for their production off these target nuclei at laboratory polar angles of 0$^{\circ}$--10$^{\circ}$, the momentum dependence of the ratio of these cross sections as well as the A-dependences of the ratios (transparency ratios) of the total cross cross sections for $\psi(2S)$ production at photon energy of 13 GeV within the different scenarios for the $\psi(2S)N$ absorption cross section. We also calculate the A-dependence of the ratio of $\psi(2S)$ and $J/\psi$ photoproduction transparency ratios at photon energies around of 11.5 GeV within the adopted scenarios for this cross section. We demonstrate that both the absolute and relative observables considered reveal a definite sensitivity to these scenarios. Therefore, the measurement of such observables in future experiments at the upgraded up to 22 GeV CEBAF facility in the near-threshold energy region might shed light both on the $\psi(2S)N$ absorption cross section and on its part associated with the nondiagonal process $\psi(2S)+N \to J/\psi+N$ at finite momenta, which are of crucial importance in understanding of charmonium production and suppression in high-energy nucleus--nucleus collisions in a search for the quark-gluon plasma.
\end{abstract}

\newpage

\section*{1. Introduction}

\hspace{1.5cm} The study of the production and suppression of charmonium states, such as the $J/\psi$ and $\psi(2S)$,
on a nuclear targets has received considerable experimental and theoretical interest in the last few decades and
remains a hot topic in high-energy proton--nucleus and nucleus--nucleus collisions, especially in conjunction with the
expected restoration of chiral symmetry in the hot and dense medium produced in these collisions which is characterized by the phase transition from composite hadrons to a quark-gluon plasma (QGP) [1--10]
\footnote{$^)$There was also a great interest in studying experimentally the charmonia production in relativistic proton--proton collisions with the main goal of providing reliable constraints on the production mechanism for these elementary systems and forming a baseline for understanding $pA$ and $A$-$A$ collisions (see, for instance, Refs. [11--14]).}$^)$.
According to [15], the charmonium
states and in particular those that are loosely bound ($J/\psi$, $\psi(2S)$) and that are produced within a QGP should dissociate in deconfined nuclear matter since $c{\bar c}$ pairs, from which they are composed, become unbound due to the effect of color Debye screening of the linear confining interaction between $c$ and ${\bar c}$ quarks.
As a result, the production of these states in heavy-ion collisions should be suppressed, which makes the $c{\bar c}$ bound states relevant probes of the formation of QGP in them. Therefore, the suppression of $J/\psi$ production in nucleus--nucleus and proton--nucleus collisions observed at SPS [16, 17], at RHIC [18, 19] and at LHC [20--22] and the suppression of the production of $\psi(2S)$ mesons relative to that of $J/\psi$ states in large collision systems [23--30] can be considered [15, 31] as an indicator of QGP formation and as a sign of their dissociation in it
\footnote{$^)$It should be mentioned that the $\psi(2S)$ meson dissociation in deconfined nuclear medium
is expected to be much easier compared to that of the $J/\psi$ due to its larger size and weaker binding.}$^)$.
On the other hand, another competing mechanisms may also contribute to the formed within a QGP charmonium abundance, including its regeneration via the coalescence of uncorrelated charm quarks and antiquarks [32--34], direct
charmonium production in hard parton--parton collisions and its subsequent scattering and absorption by the initial nucleons, cold nuclear matter (CNM) effects, in particular, such as its final-state interactions with comoving particles (with light mesons in the hadron gas) formed at the late stages of heavy-ion collisions when a QGP expands, cools and hadronizes [2, 31, 35]. These interactions could distort the information coming from the QGP or even wash it away. Thus, it is of great importance for the diagnostics of a QGP formation in high-energy heavy-ion collisions to establish the approximate size of the relatively low-energy
\footnote{$^)$Since the relative motion between the comoving charmonium and nuclear matter is rather slow [36].}$^)$
$c{\bar c}$ + light hadron/nucleon cross sections [2, 31, 37].
Since we have no charmonium beams or targets, the ${J/\psi}N$ "experimental" cross sections are usually inferred indirectly from ${\gamma}p \to {J/\psi}p$ data assuming the vector meson dominance hypothesis [2, 38], or from data on $J/\psi$ nuclear attenuation in ${\gamma}A$ [39, 40] and $pA$ [41, 42] reactions. Since the mid 1970s, the charmonium--nucleon interaction has been intensely debated. However, a compelling understanding of this interaction is still lacking.
The $J/\psi$ absorption cross sections in nuclear matter deduced from different experiments cluster around a value of
about 3--7 mb. Thus, Huefner and Kopeliovich [38] have used $J/\psi$ photoproduction data on the nucleon combined with a vector meson dominance model extended to a multichannel case to estimate the ${J/\psi}N$ total cross section and find a value of about 2.8(0.12)--4.1(0.15) mb for this cross section at $\sqrt{s}=10$ GeV.
The value of the $J/\psi$--nucleon absorption cross section $\sigma_{{J/\psi}N}=$ 3.5 mb was extracted from the measured $A$-dependence of $J/\psi$ absorption in the SLAC photoproduction experiment [39, 40] at photon energies $\sim$ 20 GeV.
An analysis [41, 42] of available data for proton--nucleus collisions at moderate energies ($\sqrt{s_{NN}} \approx 20$ GeV) leads to a value of about 6--7 mb of the $J/\psi$ absorption cross section on a nucleon, which is larger than that deduced in the SLAC experiment [39, 40] by a factor $\approx$ 2. The large ${J/\psi}N$ inelastic cross section in the range of 6--8 mb has been estimated in Ref. [37] adopting the effective Lagrangians. On the contrary, the authors of Ref. [43] found this cross section to be of $\approx$ 3--4 mb using the generalized vector meson dominance model.
For an overview of the subject and further references see Refs. [1, 2, 36, 37]. On the other hand, information on the
$\psi(2S)$--nucleon interaction is scarce in the literature and the total, inelastic and elastic cross sections for this interaction are poorly known. Thus, for example, the expectation based on QCD leads to the conclusion that the
$\psi(2S)N$ total cross section is  a factor of 2--4 larger than the ${J/\psi}N$ one due to the larger size of the
$\psi(2S)$ compared to that of the $J/\psi$ [38] and it may reach the values about of 20 mb [44]. For the sake of conciseness, we give more detailed information on this issue herein below.

To get a robust enough experimental information about the low-energy $\psi(2S)$ meson--nucleon interaction in cold nuclear matter, it is crucial to investigate the $\psi(2S)$ photoproduction on protons and nuclei at near-threshold energies. The advantage of photoproduction on nuclei at these energies compared to the high-energy hadronic collisions is that the interpretation of data from photoproduction experiments is clearer due to insignificant strength of initial-state photon interaction, a fewer individual exclusive elementary $\psi(2S)$ meson production channels need to be accounted for in it and since in elementary low-energy photon-induced reactions, contrary to the high-energy ones, the $\psi(2S)$ mesons are produced with relatively low momenta in the target nucleus rest-frame at which the formation time effects play inessential role and the final $\psi(2S)$s interact with nuclear matter, but not the embryonic $c{\bar c}$ ones. At present there are no measurements of the low-energy $\psi(2S)$--nucleon interaction cross sections in near-threshold photonuclear reactions. Such measurements are planned to be performed at the proposed CEBAF upgraded facility with a 22 GeV photon beam [45--48]
\footnote{$^)$It should be also mentioned that the near-threshold photoproduction of $J/\psi$ on the proton has been recently studied by the GlueX [49, 50] and $J/\psi$--007 [51] experiments at the JLab. Moreover, the first measurement of $J/\psi$ photoproduction off deuterium, helium and carbon target nuclei, using the GlueX detector at the JLab, in the photon energy range of 7 to 10.8 GeV, extending below and above the photoproduction threshold on the free proton of $\approx$ 8.2 GeV, has been reported in the very recent publication [52].}$^)$.

To motivate them,
in this work we present the detailed predictions for the absolute and relative excitation functions for production
of $\psi(2S)$ mesons off $^{12}$C and $^{184}$W target nuclei, for their absolute momentum distributions from these nuclei as well as for the A and momentum dependences of the relative cross sections for $\psi(2S)$ and $\psi(2S)$ and $J/\psi$ production from ${\gamma}A$ reactions at threshold energies obtained on the basis of the first collision model within the different plausible scenarios for the $\psi(2S)$ in-medium absorption cross section. They can be tested by future measurements at the upgraded up to 22 GeV CEBAF facility with the aim of discriminating between these scenarios.
This would clearly be of great importance both for the relevance to QGP searches and as a valuable test of the theoretical predictions in the field of charmonia scattering.

\section*{2. The model: direct $\psi(2S)$ photoproduction mechanism}

\hspace{1.5cm} Direct $\psi(2S)$ photoproduction on nuclear targets in the near-threshold laboratory incident photon
beam energy region $E_{\gamma} \le 16.4$ GeV of interest
\footnote{$^)$Which corresponds to the center-of-mass energies $W$ of the photon-proton system $W \le 5.6$ GeV,
or to the relatively "low" excess energies $\epsilon_{{\psi(2S)}p}$ above the $\psi(2S)p$ production threshold
$W_{\rm th}=\sqrt{s_{\rm th}}=m_{\psi(2S)}+m_{p}=$ 4.6244 GeV ($m_{\psi(2S)}$ and $m_{p}$ are $\psi(2S)$ meson and proton free space masses, respectively ) $0 \le \epsilon \le 1.0$ GeV and where the $\psi(2S)$ mesons can be observed in the ${\gamma}p$ and ${\gamma}A$ reactions at the JLab upgraded facility with a 22 GeV photon beam [45--48].}$^)$
may proceed via the following elementary processes with the lowest free production threshold ($\approx$ 10.93 GeV) [53]:
\begin{equation}
{\gamma}+p \to \psi(2S)+p,
\end{equation}
\begin{equation}
{\gamma}+n \to \psi(2S)+n.
\end{equation}
The $\psi(2S)$ mesons and nucleons, produced in these processes, are sufficiently energetic.
Thus, for example, the kinematically allowed $\psi(2S)$ meson and final
proton laboratory momenta in the direct process (1), proceeding on the free target proton at rest, vary within the momentum ranges of 7.478--12.058 GeV/c and 0.942--5.522 GeV/c, respectively, at incident photon energy of
$E_{\gamma}=13$ GeV. Since the neutron mass is approximately equal to the proton mass,
the kinematical characteristics of final particles, produced in the reaction (2), are similar to those of
final particles in the process (1). Since the medium effects are expected to be reduced for high momenta,
we will ignore the medium modifications of the outgoing high-momentum $\psi(2S)$ mesons and nucleons
in the case when the reactions (1), (2) proceed on a nucleons embedded in a nuclear target
\footnote{$^)$It is expected [54] that even at low energies the $\psi(2S)$ in-medium mass shift at normal nuclear
matter density is only of the order of a few MeV due to a small coupling of the $c$, ${\bar c}$ quarks to the nuclear medium.}$^)$.

  Then, assuming that at energies of interest the incident photon can reach every point inside the target nucleus
without being absorbed, instantaneous production of the full-sized $\psi(2S)$ meson on a bound nucleon at this point and describing its "normal" absorption by intranuclear nucleons by the (effective) absorption cross section $\sigma_{\psi(2S)N}$
\footnote{$^)$Which will be defined below.}$^)$, we can represent the total
\footnote{$^)$In the full phase space without any cuts on angle and momentum of the oberved $\psi(2S)$ meson.}$^)$
cross section for the production of $\psi(2S)$ mesons on nuclei from the direct photon--induced reaction channels (1), (2) as follows [55]:
\begin{equation}
\sigma_{{\gamma}A\to \psi(2S)X}^{({\rm dir})}(E_{\gamma})=I_{V}[A,\sigma_{\psi(2S)N}]
\left<\sigma_{{\gamma}p \to \psi(2S)p}(E_{\gamma})\right>_A,
\end{equation}
where the effective number of target nucleons participating in the direct ${\gamma}N \to \psi(2S)N$
processes, $I_{V}[A,\sigma_{\psi(2S)N}]$, and "in-medium" total cross section for the production of $\psi(2S)$ mesons
in reaction (1) $\sigma_{{\gamma}p \to \psi(2S)p}(\sqrt{s^*})$ at the in-medium ${\gamma}p$ center-of-mass energy $\sqrt{s^*}$, averaged over target nucleon binding and Fermi motion, $\left<\sigma_{{\gamma}p \to \psi(2S)p}(E_{\gamma})\right>_A$, are defined by Eqs. (4), (5) and (6) from Ref. [55], respectively, in which one needs to make the substitution: $X(3872) \to \psi(2S)$
\footnote{$^)$In Eq. (3) we assume that the $\psi(2S)$ meson production
cross sections in ${\gamma}p$ and ${\gamma}n$ interactions are the same.}$^)$.

  As before in Ref. [55], we suggest that the "in-medium" cross section
$\sigma_{{\gamma}p \to \psi(2S)p}({\sqrt{s^*}})$ for $\psi(2S)$ production in process (1)
is equivalent to the vacuum cross section $\sigma_{{\gamma}p \to \psi(2S)p}({\sqrt{s}})$,
in which the free space center-of-mass energy squared $s$ for given photon laboratory energy $E_{\gamma}$,
presented by the formula
\begin{equation}
s=s(E_{\gamma})=W^2=(E_{\gamma}+m_p)^2-{\bf p}_{\gamma}^2=m_p^2+2m_pE_{\gamma},
\end{equation}
is replaced by the in-medium expression
\begin{equation}
  s^*=(E_{\gamma}+E_t)^2-({\bf p}_{\gamma}+{\bf p}_t)^2.
\end{equation}
Here, ${\bf p}_{\gamma}$ is the momentum of the incident photon, $E_t$ and ${\bf p}_{t}$ are the total energy and
momentum of the struck target proton involved in the collision process (1).
The quantity $E_t$ is determined by Eq. (8) in Ref. [55].
\begin{figure}[htb]
\begin{center}
\includegraphics[width=16.0cm]{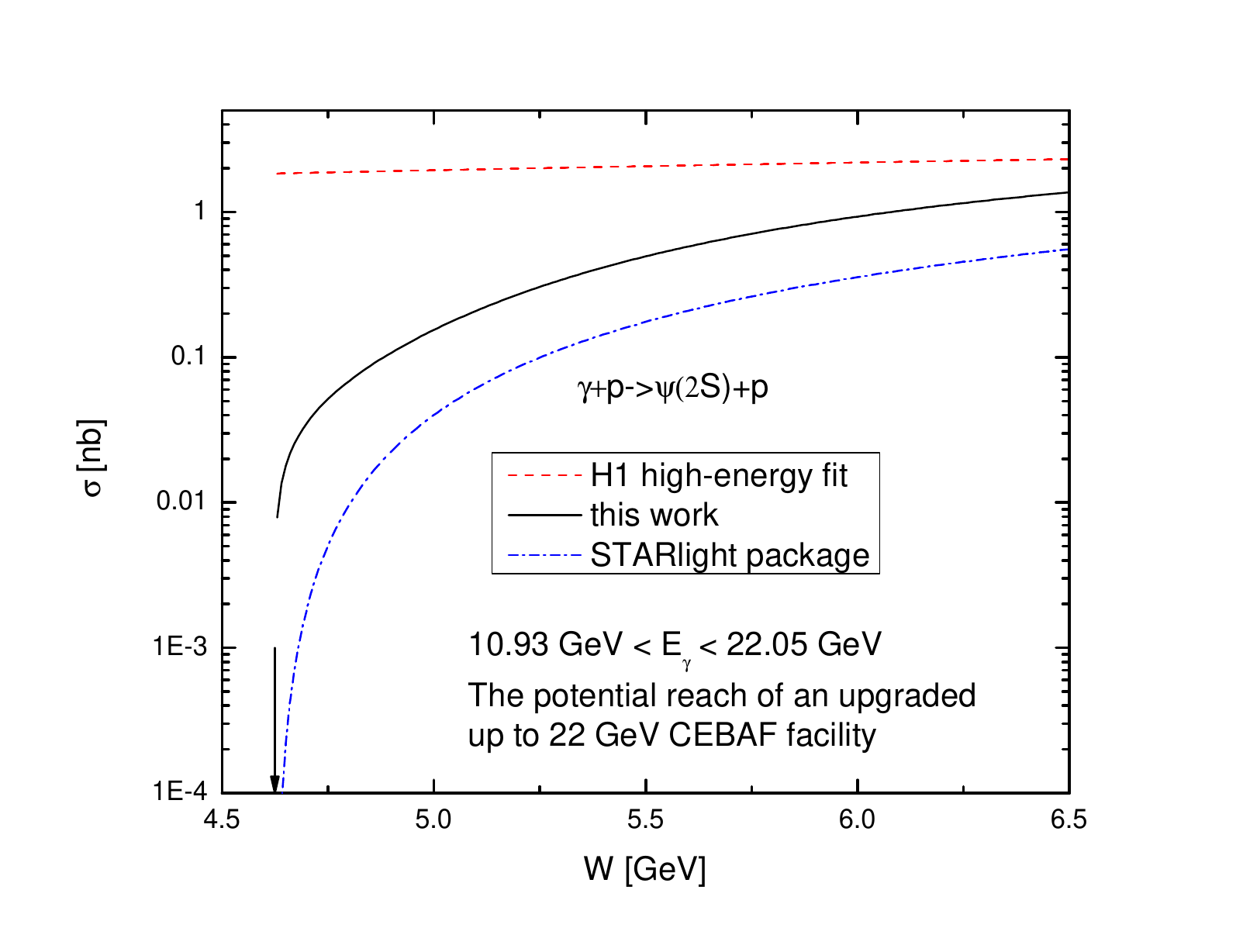}
\vspace*{-2mm} \caption{(Color online.) Total cross section for the reaction
${\gamma}p \to {\psi(2S)}p$ as a function of the center-of-mass energy $W=\sqrt{s}$ of the
photon--proton collisions in the kinematic range accessible at the proposed JLab upgraded facility
with a 22 GeV photon beam [45--48]. Solid, dashed and dotted-dashed curves represent calculations performed
using Eqs. (7)--(10), (11) and (12), respectively. The arrow indicates the threshold energy of 10.93 GeV for $\psi(2S)$
photoproduction on a free target proton at rest.}
\label{void}
\end{center}
\end{figure}
For the free total cross section $\sigma_{{\gamma}p \to \psi(2S)p}({\sqrt{s}})$ no data are available
presently at threshold energies $W \le 6.5$ GeV relevant for future CEBAF facility upgraded to the energy of 22 GeV.
The experimental data on $\psi(2S)$ meson production in the reaction ${\gamma}p \to \psi(2S)p$ are available only
at high photon-proton center-of-mass energies $W=\sqrt{s} >$ 30 GeV [56--59]. To access the total cross section of
this reaction at near-threshold photon energies, we follow a procedure, employed in the literature, see e.g. [60].
An analysis of the data on the production of $\psi(2S)$ and $J/\psi$ mesons in ${\gamma}p$ collisions in the
kinematic range of $40 < W < 150$ GeV, collected by the H1 Collaboration at the HERA $ep$ collider [56], gave the following ratio of the $\psi(2S)$ to $J/\psi$ total diffractive photoproduction cross sections in this range:
\begin{equation}
 \sigma_{{\gamma}p \to {\psi(2S)}p}(W)/\sigma_{{\gamma}p \to {J/\psi}p}(W) \approx 0.166,
\end{equation}
which is in good agreement with both an earlier measurements [57] and with more later on ones [59] in the kinematic
range of $30 < W < 180$ GeV.
Accounting for the commonality in the $J/\psi$ and $\psi(2S)$ production in ${\gamma}p$ interactions
[56], we assume that in the threshold region $W \le 6.5$ GeV the ratio of the total cross
sections of the reactions ${\gamma}p \to {\psi(2S)}p$ and ${\gamma}p \to {J/\psi}p$
is the same as that expressed by Eq. (6) derived at the same high ${\gamma}p$ c.m.s. energies. However, in this
ratio the former and latter cross sections are calculated,
respectively, at the collisional energies $\sqrt{s}$ and $\sqrt{{\tilde s}}$, which correspond to the
same excess energies $\epsilon_{{\psi(2S)}p}$ and $\epsilon_{{J/\psi}p}$ above the
${\psi(2S)}p$  and ${J/\psi}p$ thresholds, namely,
\begin{equation}
 \sigma_{{\gamma}p \to {\psi(2S)}p}(\sqrt{s})/
 \sigma_{{\gamma}p \to {J/\psi}p}(\sqrt{{\tilde s}}) \approx 0.166,
\end{equation}
where, according to the preceding, the center-of-mass energies $\sqrt{s}$ and $\sqrt{{\tilde s}}$
are linked by the relation:
\begin{equation}
\epsilon_{{J/\psi}p}=\sqrt{{\tilde s}}-\sqrt{{\tilde s}_{\rm th}}=
\epsilon_{{\psi(2S)}p}=\sqrt{s}-\sqrt{s_{\rm th}}.
\end{equation}
Here, $\sqrt{{\tilde s}_{\rm th}}=m_{J/\psi}+m_p$ ($m_{J/\psi}$ is the bare $J/\psi$ meson mass).
Thus, we have:
\begin{equation}
\sqrt{{\tilde s}}=\sqrt{s}-\sqrt{s_{\rm th}}+\sqrt{{\tilde s}_{\rm th}}=\sqrt{s}-m_{\psi(2S)}+m_{J/\psi}.
\end{equation}
Evidently, that at high energies such that $\sqrt{s} >> \sqrt{s_{\rm th}}$,
$\sqrt{{\tilde s}}$ $\approx$ $\sqrt{s}$ and the expression (7) transforms to (6). At low photon
energies $\sqrt{s} \le 6.5$ GeV of interest, the c.m.s. energy $\sqrt{{\tilde s}} \le 5.91$ GeV.
The latter corresponds to the laboratory photon energy domain $\le$ 18.15 GeV.
For the free total cross section $\sigma_{{\gamma}p \to {J/\psi}p}({\sqrt{{\tilde s}}})$
in this domain we have used the following parametrization [61] of the available here experimental data [49],
based on the predictions of the two gluon and three gluon exchange model [62] near threshold:
\begin{equation}
\sigma_{{\gamma}p \to {J/\psi}p}({\sqrt{{\tilde s}}})= \sigma_{2g}({\sqrt{{\tilde s}}})+
\sigma_{3g}({\sqrt{{\tilde s}}}),
\end{equation}
where 2$g$ and 3$g$ exchanges cross sections $\sigma_{2g}({\sqrt{{\tilde s}}})$ and
$\sigma_{3g}({\sqrt{{\tilde s}}})$ are given in Ref. [61] by formulas (7) and (8), respectively.
The results of calculations by Eqs. (7)--(10) of the diffractive total photoproduction cross section of the reaction
${\gamma}p \to {\psi(2S)}p$ at relative "low" c.m. energies $W \le 6.5$ GeV, corresponding to the energy
coverage of the upgraded up to 22 GeV CEBAF facility [45--48], are shown in Fig. 1 (solid curve).
The results from the extrapolation of the power-law fit (cf. Eq. (6) and Refs. [58, 63])
$$
\sigma_{{\gamma}p \to {\psi(2S)}p}(\sqrt{s})=0.166\cdot\sigma_{{\gamma}p \to {J/\psi}p}(\sqrt{s}),
$$
\begin{equation}
\sigma_{{\gamma}p \to {J/\psi}p}(\sqrt{s})=81\left(\frac{\sqrt{s}}{90~{\rm GeV}}\right)^{0.67}~[\rm nb]
\end{equation}
of the high-energy data on the ${\gamma}p \to {\psi(2S)}p$ and ${\gamma}p \to {J/\psi}p$ cross sections, obtained
in the H1 analysis [56, 64], to the threshold energies are shown in Fig. 1 as well (dashed curve).
In addition, the predictions from the current parametrization
\begin{equation}
 \sigma_{{\gamma}p \to {\psi(2S)}p}(\sqrt{s})=0.674\left(1-\frac{s_{\rm th}}{s}\right)^2\cdot(\sqrt{s})^{0.65}~[\rm nb],
\end{equation}
adopted in the STARlight Monte Carlo simulation program package [65] to simulate the $\psi(2S)$ production in ultra-peripheral collisions of relativistic ions, are also shown in Fig. 1 (dotted-dashed curve).
This figure shows that our parametrization (7)--(10) (solid curve) deviates substantially from
the upper-lying and lower-lying curves corresponding to the calculations in line with
Eqs. (11) and (12), respectively, at all energies considered, especially, for energies close to the threshold.
In particular, at photon energies around 5.0 GeV our parametrization (7)--(10) is larger (by factor of about 4) than the results obtained using the parametrization (12), adopted in the STARlight package.
On the other hand, here it is significantly smaller (by factor of about 12) than the results obtained from the high-energy fit (11). The total cross section of the ${\gamma}p \to {\psi(2S)}p$ reaction, predicted by us, is of
the order of 0.1--1 nanobarns for photon energies $W\sim$ 5.0--6.5 GeV, which are in reasonable agreement with the
model prediction [53] in this energy regime and which together with this prediction could be measured in the future
at the CEBAF facility upgraded to the energy of 22 GeV.
Therefore, in our subsequent calculations we will use the empirical formulas (7)--(10) as a guideline for
a reasonable estimation of the $\psi(2S)$ yield in ${\gamma}A$ reactions.

We focus now on the local proton and neutron densities, adopted in the calculations of the quantity $I_{V}[A,\sigma_{\psi(2S)N}]$, entering into Eq. (3), for the target nuclei $^{12}_{6}$C, $^{27}_{13}$Al, $^{40}_{20}$Ca, $^{63}_{29}$Cu, $^{93}_{41}$Nb, $^{112}_{50}$Sn, $^{184}_{74}$W, $^{208}_{82}$Pb and $^{238}_{92}$U considered in the present work. They are given in Ref. [55]. As in Ref. [55], for medium-weight $^{93}_{41}$Nb, $^{112}_{50}$Sn and heavy $^{184}_{74}$W, $^{208}_{82}$Pb, $^{238}_{92}$U target nuclei we use the neutron density $\rho_n(r)$ in the 'skin' form.

To estimate the rate of the $\psi(2S)$ photoproduction on nuclei, we have to specify the effective input $\psi(2S)$--nucleon absorption cross section $\sigma_{{\psi(2S)}N}$, which determines the number of target nucleons participating in the direct ${\gamma}N \to \psi(2S)N$ processes, $I_{V}[A,\sigma_{\psi(2S)N}]$, (cf. Eqs. (4), (5) from Ref. [55]) and
which is not well known currently. Therefore, we have to rely on some theoretical predictions and estimates for it,
existing in the literature in this energy region. The above cross section can be evaluated as a sum of the genuine
$\psi(2S)$--nucleon inelastic cross section $\sigma_{{\psi(2S)}N}^{\rm gen}$ and
the $\psi(2S)N$ inelastic cross section $\sigma_{{\psi(2S)}N}^{\rm non}$ arising, respectively, mostly from the charm-transfer inelastic channels $\psi(2S)+N \to \Lambda_c+{\bar D}$, $\psi(2S)+N \to \Sigma_c+{\bar D}$ and from
the nondiagonal inelastic process $\psi(2S)+N \to J/\psi+N$ (the $\psi(2S)\to J/\psi$ transition) [36, 43].
As was already mentioned above, the expectation [38] based on QCD tells us that the genuine
${\psi(2S)N}$ total cross section is a factor of 2--4 larger than the genuine ${J/\psi}N$ total cross section due to
the larger characteristic size $r_{\psi(2S)}$ of the radial excitation $\psi(2S)$ compared to that $r_{J/\psi}$ of the 1$S$ $J/\psi$ ground state ($r_{\psi(2S)} \approx 2r_{J/\psi}$ [44, 66]) and it can reach the values of the order of 20 mb [44]. When this relation between the $\psi(2S)$ and $J/\psi$ total interaction cross sections is extended to their inelastic interaction cross sections [67], the value $\sigma_{{J/\psi}N} \approx$ 3.5 mb of the genuine $J/\psi$--nucleon absorption cross section extracted from the $A$-dependence of the $J/\psi$ photoproduction studied at SLAC at photon energies $\sim$ 20 GeV [39, 40] leads to the values $\sim$ 7--14 mb for the absorption cross section $\sigma_{{\psi(2S)}N}^{\rm gen}$
\footnote{$^)$It is also worth noting that due to the fact that the genuine $\psi(2S)N$ elastic cross section is expected to be small and therefore the genuine $\psi(2S)N$ total cross section is almost completely exhausted by its inelastic cross section (the ratio of the elastic to the total $\psi(2S)N$ cross section is approximately 8.5--10\%), these values are in line with those of $\sim$ 7.5--8.5 mb evaluated in Ref. [43] for the genuine $\psi(2S)$--nucleon inelastic cross section within the generalized vector meson dominance model. They are also well consistent with those of $\sim$ 8--11 mb, which can be obtained using parametrization (2.22) from Ref. [68] for the $\psi(2S)$--nucleon total cross section at relevant $\psi(2S)$ laboratory momenta (see above) and with those of $\sim$ 10 mb found in Refs. [69, 70] for the
latter cross section at $\sqrt{s_{\psi(2S)N}} \approx 10$ GeV. For a short review of the experimental situation and theoretical results for the low-energy dissociation cross sections of charmonia ($J/\psi$ and $\psi(2S)$) by light
hadrons ($\pi$, $\rho$, $\omega$) and nucleons see Ref. [2] as well. A very recent theoretical results for these cross sections are reported, in particular, in Ref. [31].}$^)$.
According to the estimates performed in Ref. [36] using the multipole expansion and
low-energy theorems in QCD, the nondiagonal $\psi(2S)N$ inelastic cross section $\sigma_{{\psi(2S)}N}^{\rm non}$ can reach tens of millibarn at rather low momenta due to its inverse-velocity behavior. But at finite momenta studied in
the present work it reaches, as is expected from Ref. [43], the values $\sim$ 0.2--0.3 mb which are less, for example, than those $\sim$ 7.5--8.5 mb evaluated in Ref. [43] for the genuine $\psi(2S)$--nucleon inelastic cross
$\sigma_{{\psi(2S)}N}^{\rm gen}$ by about of one to two orders of magnitude. This implies that the effective
$\psi(2S)$ meson--nucleon absorption cross section $\sigma_{{\psi(2S)}N}$ is almost completely governed by its
genuine cross section $\sigma_{{\psi(2S)}N}^{\rm gen}$ and the inclusive $\psi(2S)$ photoproduction experiment
will allow practically a direct measurement of the latter cross section.
In view of the above, in our present study we will adopt the following three realistic options for the effective
(and practically for the genuine) $\psi(2S)$ absorption cross section: 7, 14, and 21 mb
and additional two ones with zero and 28 mb values to extend the range of applicability of our model.

Obviously, the target nucleon binding and Fermi motion play a relatively minor role in the production of
$\psi(2S)$ mesons on nuclei at photon energies well above threshold and in the first approximation
we can ignore it here in this production. Then, from Eq. (3) we get the following simple expression for the total $\psi(2S)$ photoproduction cross section $\sigma_{{\gamma}A \to \psi(2S)X}^{({\rm dir})}(E_{\gamma})$:
\begin{equation}
\sigma_{{\gamma}A\to \psi(2S)X}^{({\rm dir})}(E_{\gamma}) \approx I_{V}[A,\sigma_{\psi(2S)N}]
\sigma_{{\gamma}p \to \psi(2S)p}(\sqrt{s(E_{\gamma})}),
\end{equation}
where the elementary cross section $\sigma_{{\gamma}p \to \psi(2S)p}(\sqrt{s(E_{\gamma})})$ is given above
by Eqs. (7)--(10).
Indeed, our calculations have shown that the difference between the expressions (3) and (13) is small at
photon beam energies around 11.5 GeV and more, and it becomes essential for photon energies close to the
threshold energy of 10.93 GeV (compare magenta short-dashed and pink short-dashed-dotted curves in figures 2
and 3 given below).
As a measure for the $\psi(2S)$ absorption cross section $\sigma_{\psi(2S)N}$ in nuclei we will use
the so-called $\psi(2S)$ transparency ratio defined as (see Ref. [55] and references herein):
\begin{equation}
S_A=\frac{\sigma_{{\gamma}A \to \psi(2S)X}(E_{\gamma})}{A~\sigma_{{\gamma}p \to \psi(2S)p}(\sqrt{s(E_{\gamma})})},
\end{equation}
{\it i.e.} the ratio of the inclusive nuclear $\psi(2S)$ photoproduction cross section divided by $A$ times the same quantity on a free proton. It is natural to assume that the direct processes (1) and (2) are dominant in the $\psi(2S)$ photoproduction on nuclei in the kinematic range of interest. Then, at photon energies well above threshold, belonging to this range, and in the case of ignoring nucleon Fermi motion and binding in the target nucleus, from Eqs. (13), (14) we have:
\begin{equation}
S_A=\frac{\sigma_{{\gamma}A \to \psi(2S)X}^{({\rm dir})}(E_{\gamma})}{A~\sigma_{{\gamma}p \to \psi(2S)p}(\sqrt{s(E_{\gamma})})} \approx \frac{I_{V}[A,\sigma_{\psi(2S)N}]}{A}.
\end{equation}
In the case of no absorption, classical probabilistic formula (15) gives, $S_A=1$, otherwise $S_A < 1$.
This simple expression allows one to extract reliable values for heavy quarkonium interaction cross sections
from photoproduction data in the considered kinematic region, since in it the effects of the coherence and
formation lengths are expected to be small [38, 44, 67].
When the target nucleon binding and Fermi motion are accounted for,
the quantity $S_A$ takes on, according to Eqs. (3) and (15), a more complex form:
\begin{equation}
S_A=\frac{I_{V}[A,\sigma_{\psi(2S)N}]}{A}\frac{\left<\sigma_{{\gamma}p \to \psi(2S)p}(E_{\gamma})\right>_A}{\sigma_{{\gamma}p \to \psi(2S)p}(\sqrt{s(E_{\gamma})})}.
\end{equation}
To avoid systematic uncertainties in the comparison with
experimental data, for instance, due to meson production in secondary processes, the transparency ratio of the
meson for a heavy target with mass number $A$ is often normalized to the transparency ratio for a light nucleus
like $^{12}$C [71]. This and Eq. (16) gives for $\psi(2S)$:
\begin{equation}
T_A=\frac{S_A}{S_C}=\frac{12~\sigma_{{\gamma}A \to \psi(2S)X}^{({\rm dir})}(E_{\gamma})}{A~\sigma_{{\gamma}{\rm C} \to \psi(2S)X}^{({\rm dir})}(E_{\gamma})}=
\frac{12~I_V[A,\sigma_{\psi(2S)N}]}{A~I_V[{\rm C},\sigma_{\psi(2S)N}]}
\frac{\left<\sigma_{{\gamma}p \to \psi(2S)p}(E_{\gamma})\right>_A}
{\left<\sigma_{{\gamma}p \to \psi(2S)p}(E_{\gamma})\right>_{\rm C}}.
\end{equation}
In the case of no nuclear effects considered and not related to the absorption of $\psi(2S)$ mesons,
the expression (17) for the quantity $T_A$ is reduced to the following simpler form:
\begin{equation}
T_A \approx \frac{12~I_V[A,\sigma_{\psi(2S)N}]}{A~I_V[{\rm C},\sigma_{\psi(2S)N}]},
\end{equation}
which gives $T_A=1$ for the no absorption case.
The formulas (13)--(18) will be used in our calculations of the above-considered integral observables.
\begin{figure}[!h]
\begin{center}
\includegraphics[width=15.0cm]{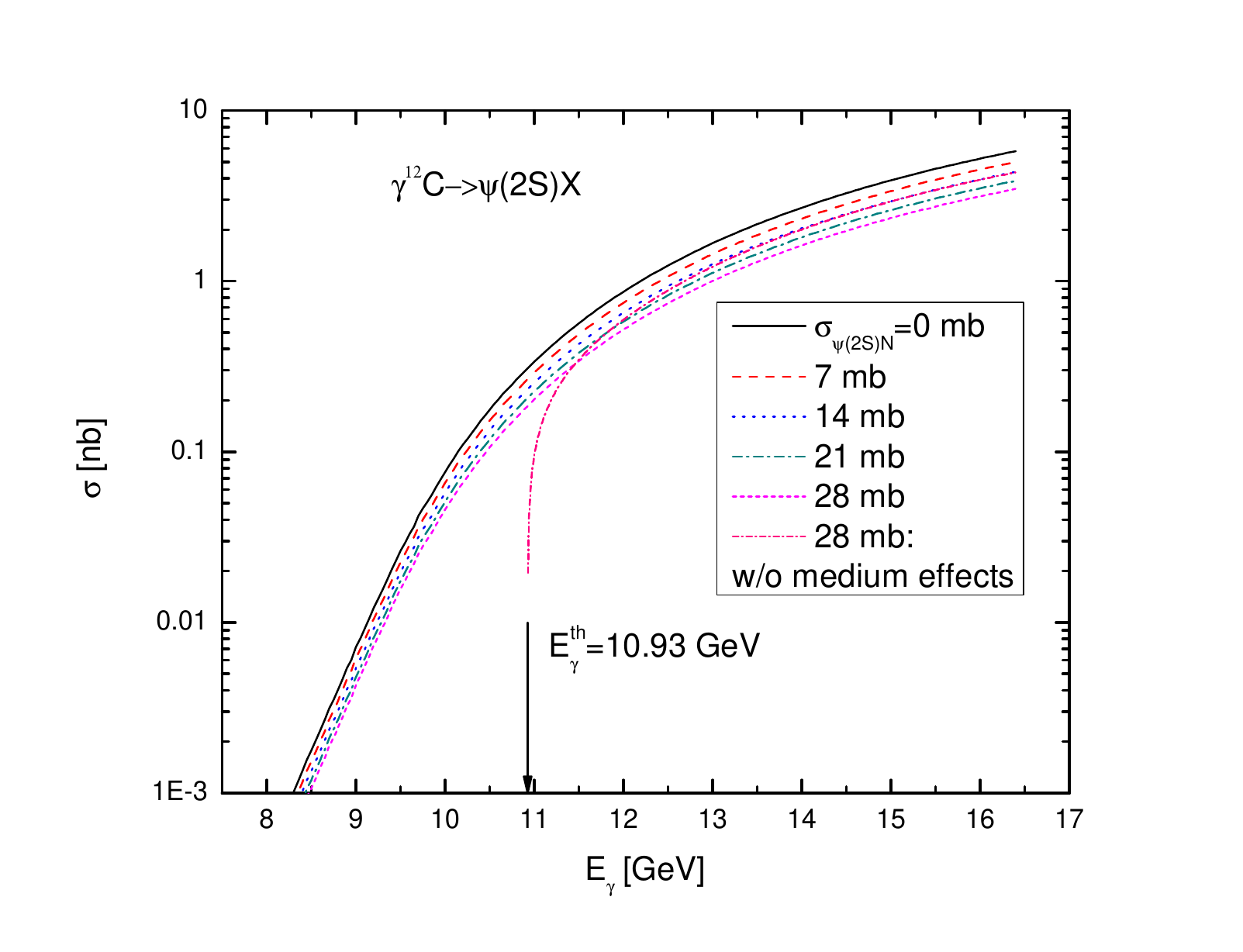}
\vspace*{-2mm} \caption{(Color online.) Excitation function for production of $\psi(2S)$
mesons off $^{12}$C from direct ${\gamma}p(n) \to {\psi(2S)}p(n)$ reactions
proceeding on an off-shell target nucleons and on a free ones being at rest. The curves are calculations for
$\sigma_{\psi(2S)N}=$ 0, 7, 14, 21 and 28 mb. The arrow indicates the threshold energy for the $\psi(2S)$
photoproduction on a free nucleon.}
\label{void}
\end{center}
\end{figure}

The information on the $\psi(2S)$ absorption cross section $\sigma_{\psi(2S)N}$ can also be extracted
from the comparison of the measured and calculated momentum distributions of $\psi(2S)$ mesons from nuclei
in the photon energy range of interest.
Therefore, we consider now the momentum-dependent inclusive differential cross section for their production
with momentum $p_{\psi(2S)}$ from the direct processes (1) and (2) in ${\gamma}A$ interactions.
Given the fact that the $\psi(2S)$ meson moves in the nucleus essentially forward in the lab frame
\footnote{$^)$Thus, for example, the maximum angle of its production on a free target proton at rest in
reaction (1) is about 5.0$^{\circ}$ at photon energy of 13 GeV.}$^)$,
we will calculate the $\psi(2S)$ momentum distribution from considered target nuclei
for the laboratory solid angle ${\Delta}{\bf \Omega}_{\psi(2S)}$ = $0^{\circ} \le \theta_{\psi(2S)} \le 10^{\circ}$,
and $0 \le \varphi_{\psi(2S)} \le 2{\pi}$. Then, according to the results presented both in Ref. [72] and above by
Eq. (3), we can obtain the following expression for this distribution:
\begin{equation}
\frac{d\sigma_{{\gamma}A\to {\psi(2S)}X}^{({\rm dir})}
(p_{\gamma},p_{\psi(2S)})}{dp_{\psi(2S)}}=
2{\pi}I_{V}[A,\sigma_{{\psi(2S)}N}]
\int\limits_{\cos10^{\circ}}^{1}d\cos{{\theta_{\psi(2S)}}}
\left<\frac{d\sigma_{{\gamma}p\to {\psi(2S)}{p}}(p_{\gamma},
p_{\psi(2S)},\theta_{\psi(2S)})}{dp_{\psi(2S)}d{\bf \Omega}_{\psi(2S)}}\right>_A,
\end{equation}
where
$\left<\frac{d\sigma_{{\gamma}p \to {\psi(2S)}p}(p_{\gamma},
p_{\psi(2S)},\theta_{\psi(2S)})}{dp_{\psi(2S)}d{\bf \Omega}_{\psi(2S)}}\right>_A$
is the off-shell differential cross section for production of $\psi(2S)$ mesons
with momentum ${\bf p}_{\psi(2S)}$ in the process ${\gamma}p \to {\psi(2S)}p$,
averaged over the Fermi motion and binding energy of the intranuclear protons.
It can be expressed by Eqs. (28), (31)--(39) from Ref. [72], in which one needs to make the
substitution: $\Upsilon(1S) \to \psi(2S)$. For better readability of this paper, we do not give
these expressions here. In order to calculate the c.m. $\psi(2S)$ angular distribution in process (1)
(cf. Eq. (34) from Ref. [72]) one needs to know its exponential $t$-slope parameter $b_{\psi(2S)}$ in
the threshold energy region. According to the estimates presented in [55],
this parameter $\approx 2.0$ GeV$^{-2}$ for incident photon energy of 13 GeV. It should be pointed out that
this value for the parameter $b_{\psi(2S)}$ agrees well with that of 1.97 GeV$^{-2}$ which can be found
from a fit to the currently available high-energy $\psi(2S)$ data at this energy, obtained by the
LHCb Collaboration in [73]. We will use it in our subsequent differential cross-section calculations.
\begin{figure}[!h]
\begin{center}
\includegraphics[width=15.0cm]{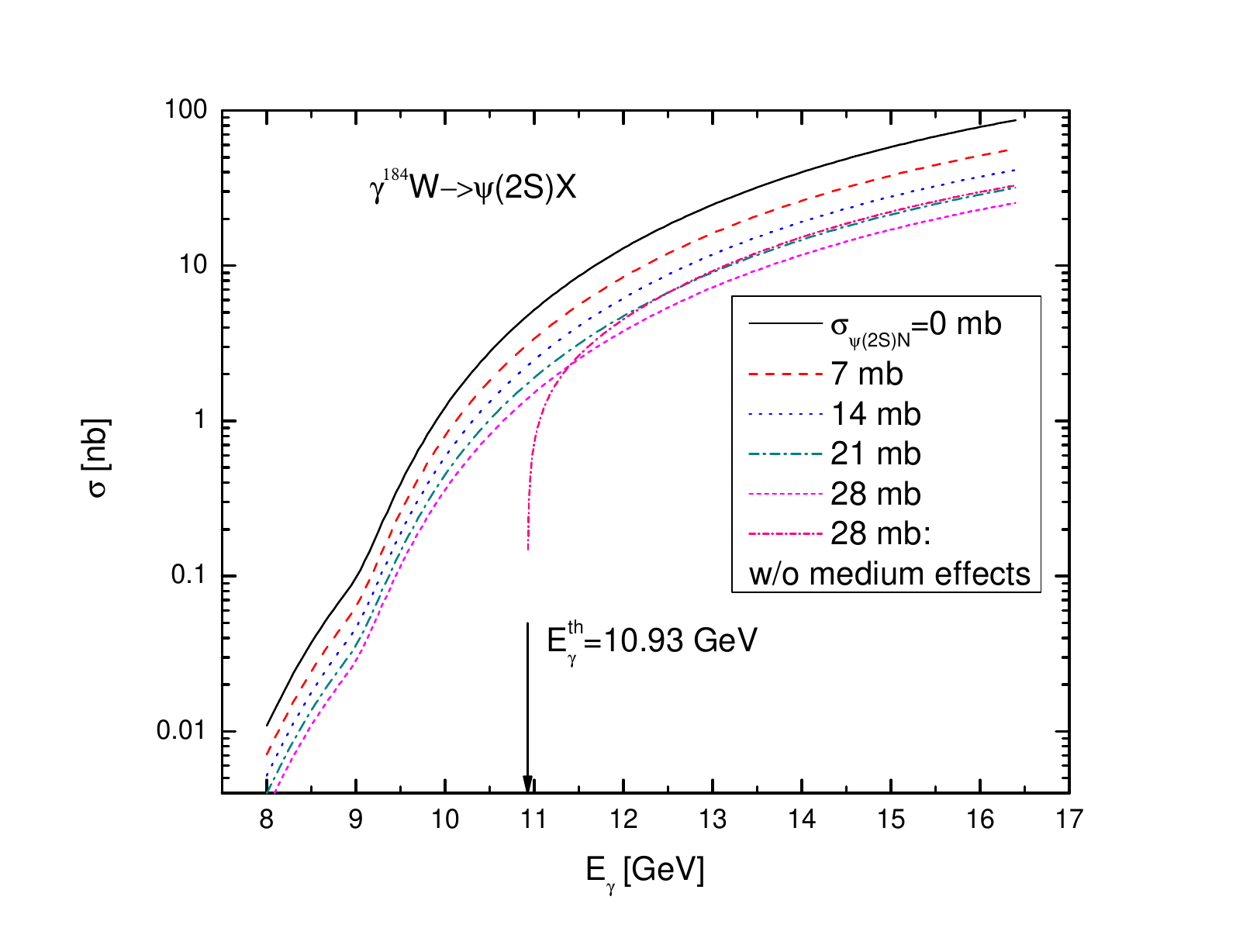}
\vspace*{-2mm} \caption{(Color online.) The same as in Fig. 2, but for the $^{184}$W target nucleus.}
\label{void}
\end{center}
\end{figure}
\begin{figure}[!h]
\begin{center}
\includegraphics[width=15.0cm]{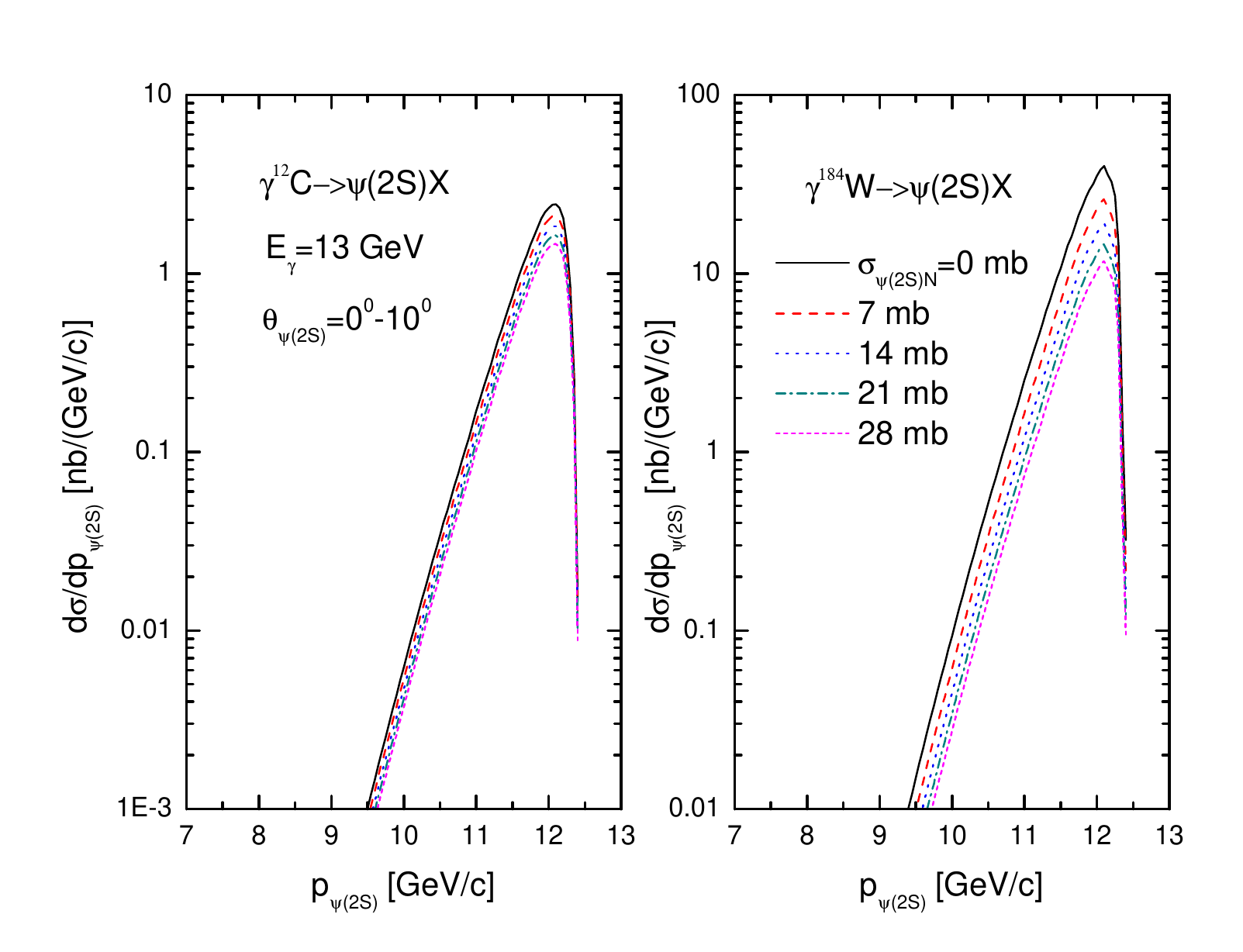}
\vspace*{-2mm} \caption{(Color online.) Momentum differential cross sections for the production of $\psi(2S)$ mesons
from the direct ${\gamma}p(n) \to {\psi(2S)}p(n)$ processes in the laboratory polar angular range of
0$^{\circ}$--10$^{\circ}$ in the interaction of photons having vacuum energy of $E_{\gamma}=$ 13 GeV with $^{12}$C
(left) and $^{184}$W (right) nuclei, calculated for different values of the $\psi(2S)$
absorption cross section $\sigma_{\psi(2S)N}$ in nuclei indicated in the inset.}
\label{void}
\end{center}
\end{figure}
\begin{figure}[!h]
\begin{center}
\includegraphics[width=15.0cm]{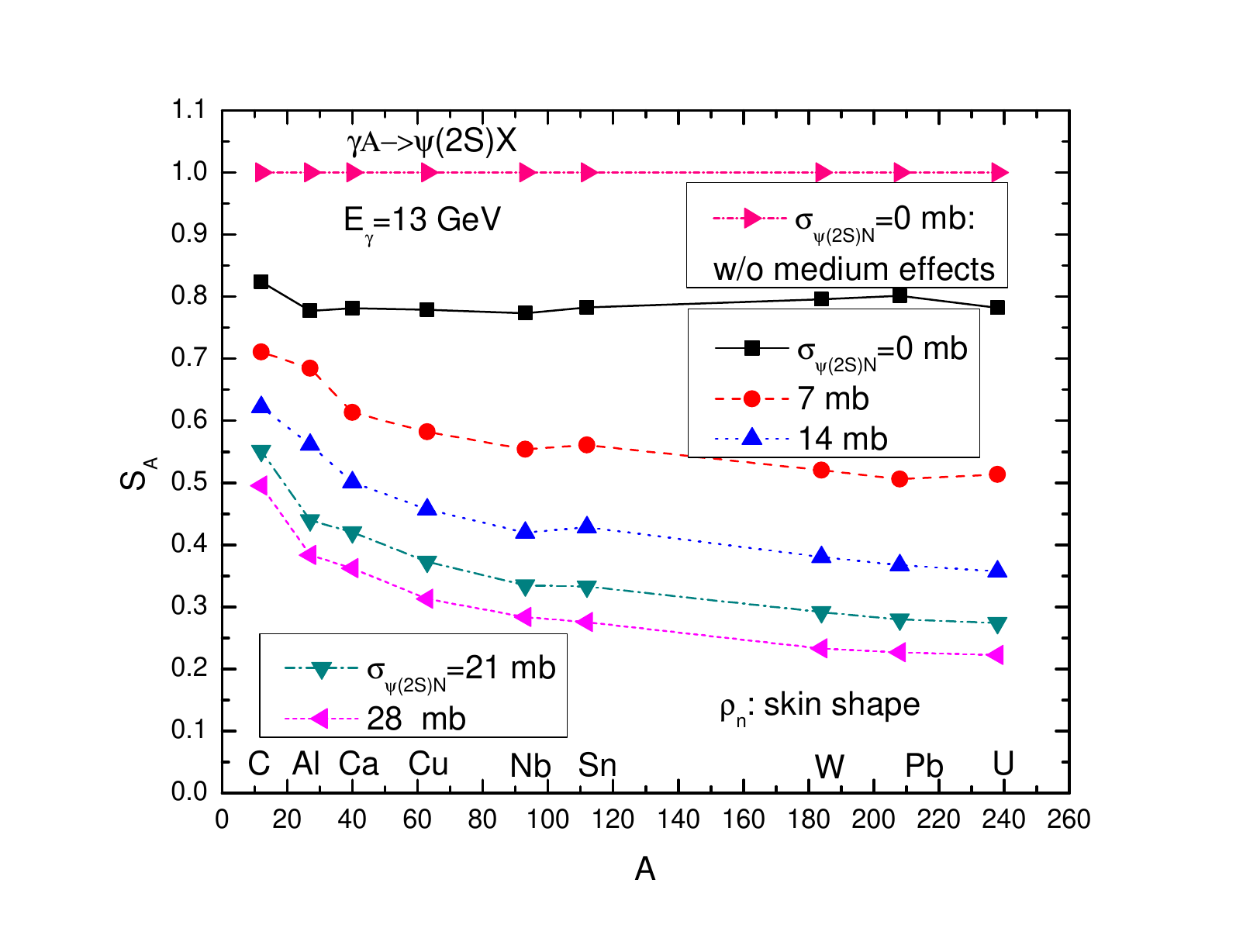}
\vspace*{-2mm} \caption{(Color online.) Transparency ratio $S_A$ for the $\psi(2S)$ mesons
from direct ${\gamma}p(n) \to {\psi(2S)}p(n)$ processes proceeding on an off-shell target nucleons
and on a free ones being at rest at incident photon energy of 13 GeV in the laboratory system as a function of the nuclear mass number $A$, calculated for different values of the $\psi(2S)$ absorption cross section $\sigma_{\psi(2S)N}$ in nuclei indicated in the insets. The lines are to guide the eyes.}
\label{void}
\end{center}
\end{figure}
\begin{figure}[!h]
\begin{center}
\includegraphics[width=15.0cm]{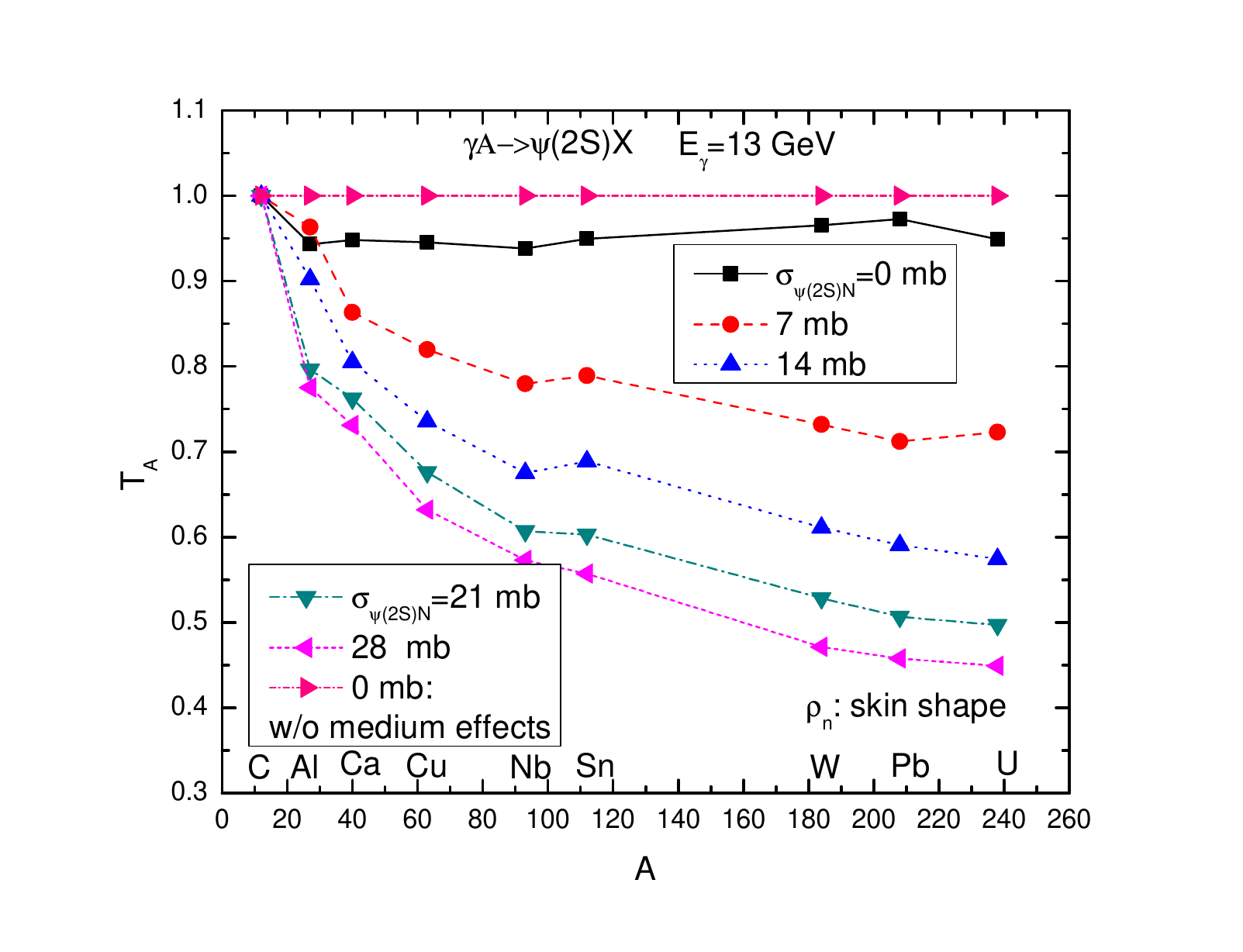}
\vspace*{-2mm} \caption{(Color online.) Transparency ratio $T_A$ for the $\psi(2S)$ mesons
from direct ${\gamma}p(n) \to \psi(2S)p(n)$ processes proceeding on an off-shell target nucleons
and on a free ones being at rest at incident photon energy of 13 GeV in the laboratory system as a function of the nuclear mass number $A$, calculated for different values of the $\psi(2S)$ absorption cross section $\sigma_{\psi(2S)N}$ in nuclei indicated in the inset. The lines are to guide the eyes.}
\label{void}
\end{center}
\end{figure}
\begin{figure}[!h]
\begin{center}
\includegraphics[width=15.0cm]{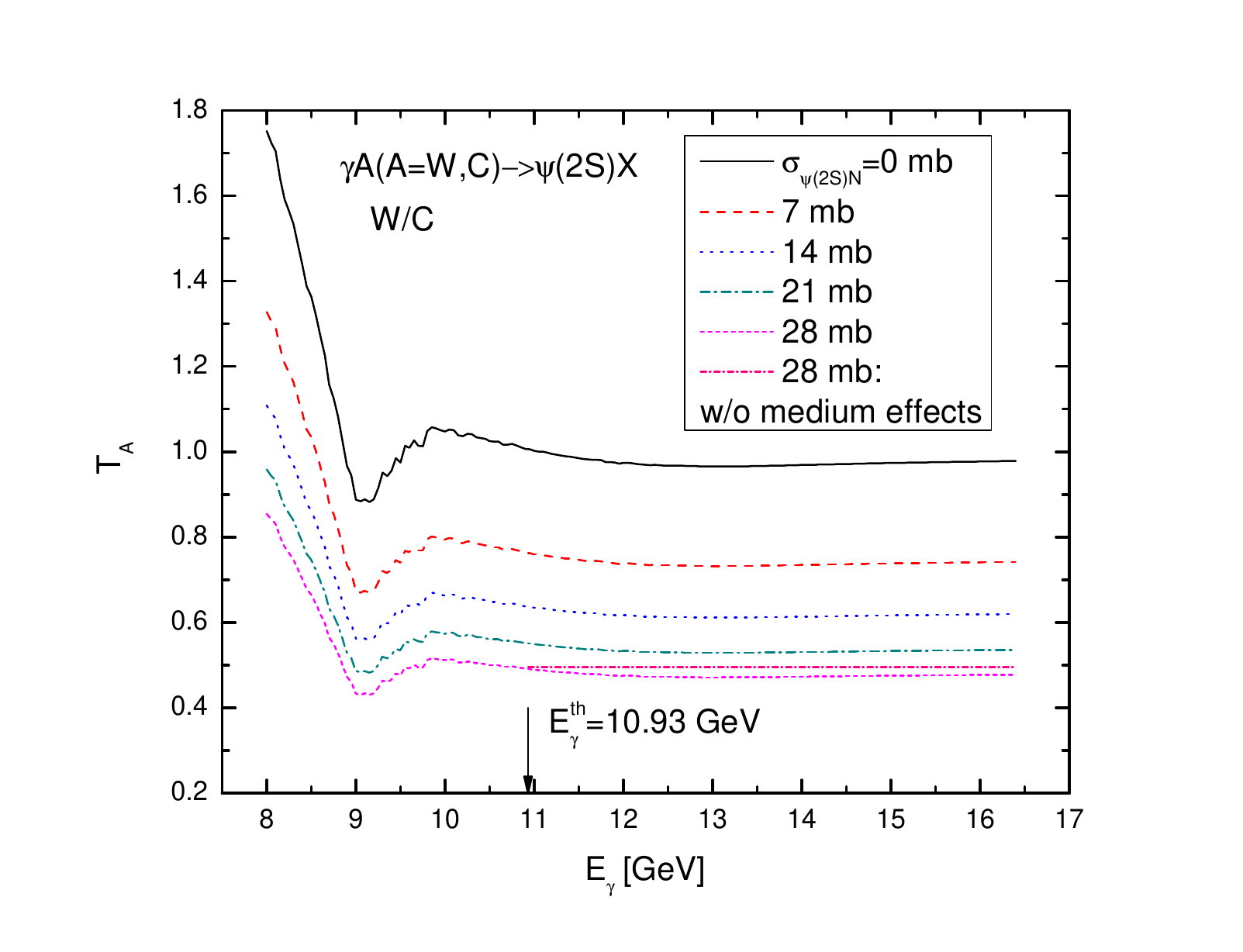}
\vspace*{-2mm} \caption{(Color online.) Transparency ratio $T_A$ for the $\psi(2S)$ mesons
from direct ${\gamma}p(n) \to {\psi(2S)}p(n)$ processes proceeding on an off-shell and free target nucleons
as a function of the incident photon energy for combination $^{184}$W/$^{12}$C, calculated for different
values of the $\psi(2S)$ absorption cross section $\sigma_{\psi(2S)N}$ in nuclei indicated in the inset.
The arrow indicates the threshold energy for the $\psi(2S)$ photoproduction on a free target
nucleon at rest.}
\label{void}
\end{center}
\end{figure}
\begin{figure}[!h]
\begin{center}
\includegraphics[width=15.0cm]{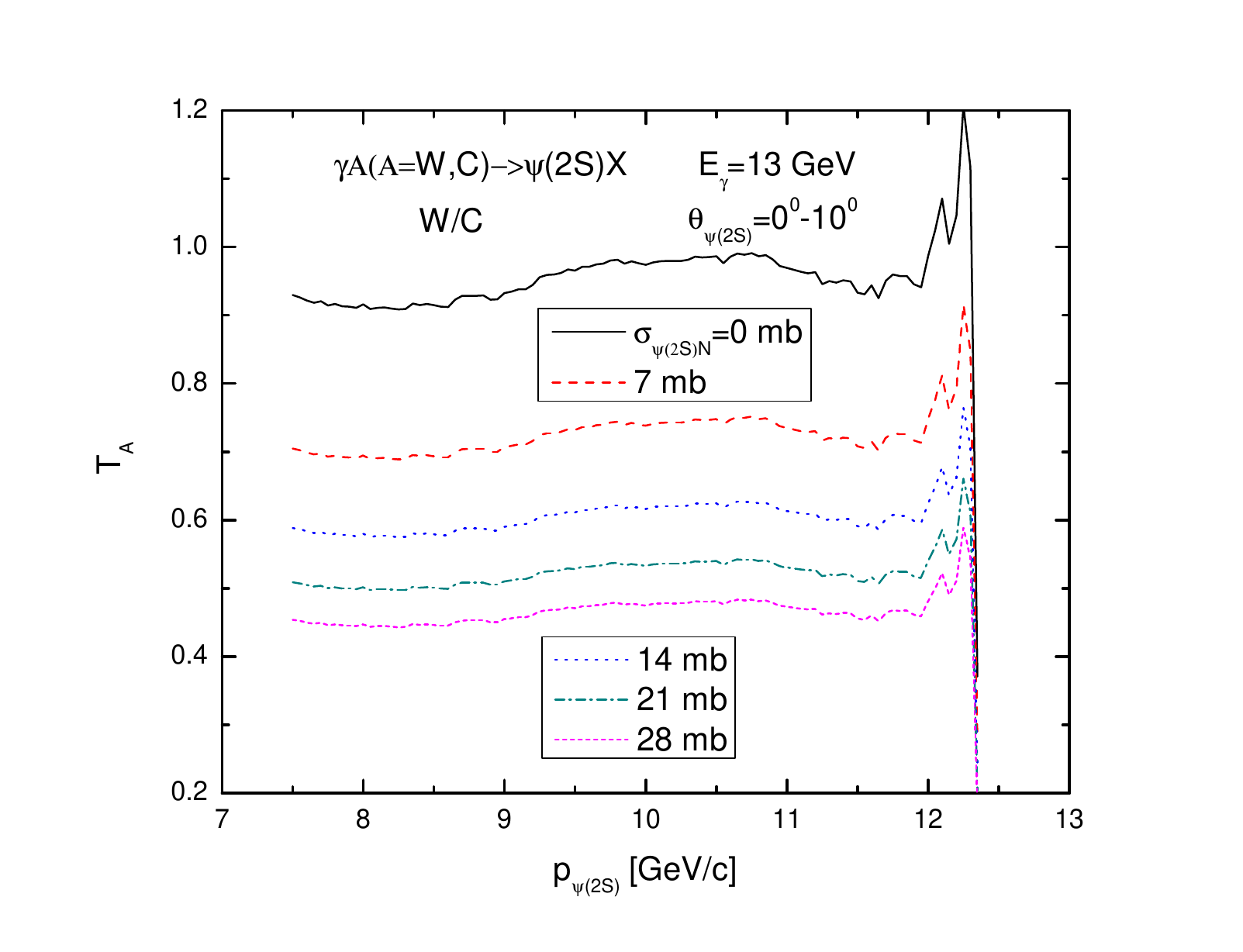}
\vspace*{-2mm} \caption{(Color online.) Transparency ratio $T_A$ for the $\psi(2S)$ mesons
from direct ${\gamma}p(n) \to \psi(2S)p(n)$ processes proceeding on an off-shell target nucleons
as a function of the $\psi(2S)$ laboratory momentum for incident photon energy of 13 GeV for combination $^{184}$W/$^{12}$C, calculated in the laboratory polar angular range of 0$^{\circ}$--10$^{\circ}$
for different values of the $\psi(2S)$ absorption cross section $\sigma_{\psi(2S)N}$ in nuclei indicated in the insets.}
\label{void}
\end{center}
\end{figure}
\begin{figure}[!h]
\begin{center}
\includegraphics[width=15.0cm]{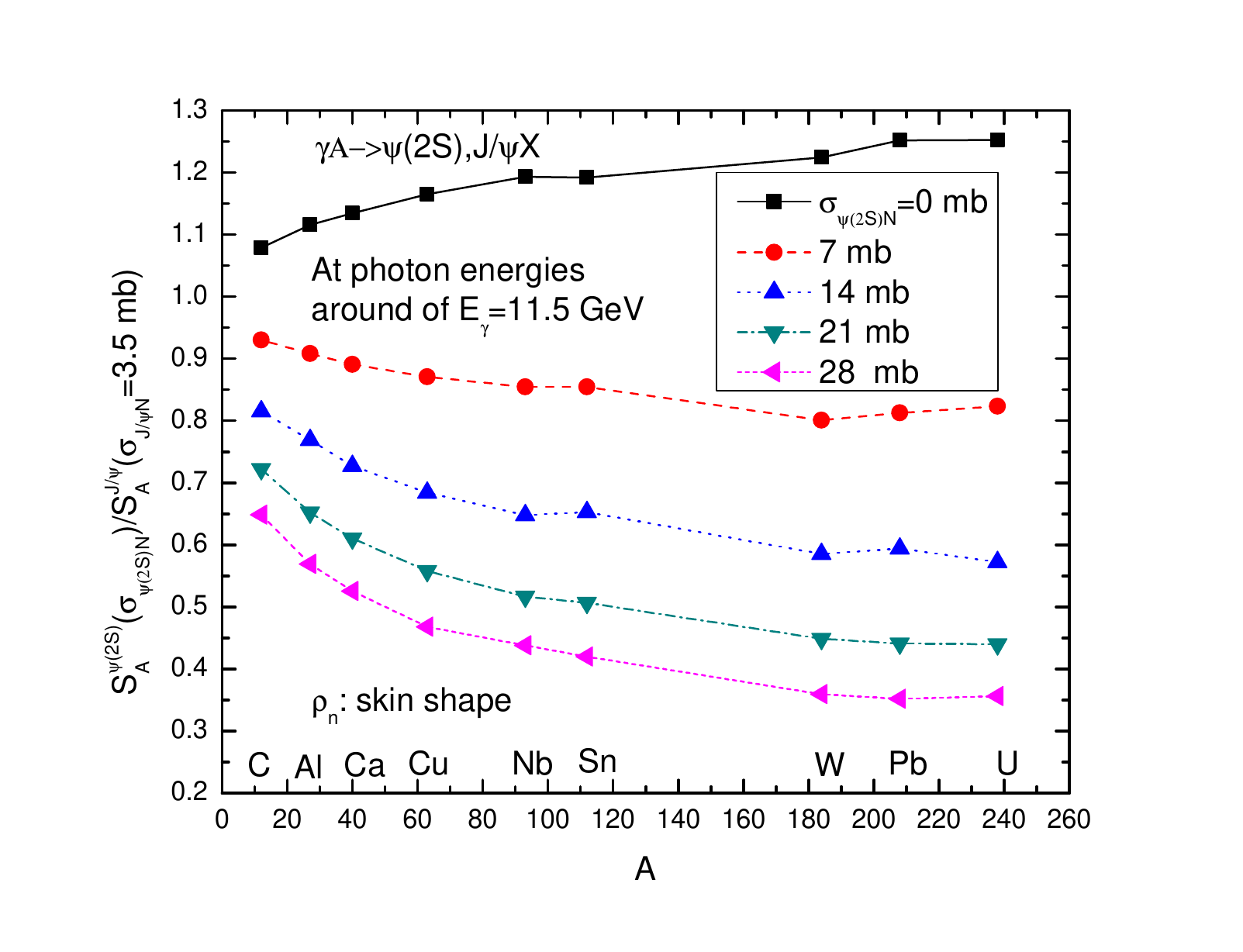}
\vspace*{-2mm} \caption{(Color online.) Nuclear mass number $A$-dependence of the
ratio of $\psi(2S)$ and $J/\psi$ photoproduction transparency ratios for incident photon energies around of 11.5 GeV,
calculated for different values of the $\psi(2S)$ absorption cross section $\sigma_{\psi(2S)N}$ in nuclei indicated in the inset and for the value of the $J/\psi$--nucleon absorption cross section $\sigma_{{J/\psi}N}=$ 3.5 mb motivated by the results from the $J/\psi$ photoproduction experiment at SLAC [39, 40].}
\label{void}
\end{center}
\end{figure}

\section*{3. Numerical results and discussion}

\hspace{1.5cm} The excitation functions for production of $\psi(2S)$ mesons off $^{12}$C and $^{184}$W nuclei,
calculated in line with Eq. (3) for five adopted options for the $\psi(2S)$ absorption cross section $\sigma_{\psi(2S)N}$ in nuclear medium as well as in line with formula (13) for the free target nucleon at rest with the value of $\sigma_{\psi(2S)N}=28$ mb, are shown in Figs. 2 and 3, respectively.
It is seen that the difference between calculations with and without taking into account
the target nucleon Fermi motion (between magenta short-dashed and pink short-dashed-dotted curves)
is sufficiently small at well above threshold photon beam energies $\sim$ 11.5--16.4 GeV,
while at lower incident energies its influence on the $\psi(2S)$ yield is significant.
It is also seen yet that for the heavy target nucleus $^{184}$W the obtained results depend strongly on the $\psi(2S)$--nucleon absorption cross section. We observe here a well distinguishable and experimentally measurable
differences $\sim$ 25--50\% between excitation functions calculated with different values of the $\psi(2S)N$ absorption
cross section. For the light target nucleus $^{12}$C, the sensitivity of the cross sections to these values becomes lower and, respectively, the same differences as above become somewhat smaller. They are $\sim$ 12--16\% and will probably be
experimentally accessible as well in the future dedicated experiments at JLab, if the respective measurements will be performed with precision  better than 6--8\%. Such measurements look quite promising since the absolute values of
the $\psi(2S)$ meson total photoproduction cross sections have a measurable strength $\sim$ 0.3--6.0 nb and 3--90 nb for carbon and tungsten target nuclei, correspondingly, at above threshold photon energies of $\sim$ 11.5--16.4 GeV.
To motivate such measurements at JLab, it is desirable to estimate the $\psi(2S)$ production rates (the event numbers) in the ${\gamma}^{12}$C and ${\gamma}^{184}$W reactions
\footnote{$^)$The $\psi(2S)$ mesons could be reconstructed via their decays $\psi(2S) \to \mu^+\mu^-$ into the dimuon final state [35, 56, 73] with branching ratio of about 0.77\% [56].}$^)$. For this purpose, we translate the $\psi(2S)$ photoproduction total cross sections, indicated above, into the expected event numbers for the $\psi(2S)$ from the reactions ${\gamma}^{12}{\rm C}(^{184}{\rm W}) \to \psi(2S)X$, $\psi(2S) \to \mu^+\mu^-$.
To estimate the total numbers of the $\psi(2S)$ events in a one-year run at the CEBAF facility, one needs
to multiply the above $\psi(2S)$ photoproduction total cross sections of 0.3--6.0 nb and 3--90 nb
on the carbon and tungsten target nuclei, respectively, by the integrated luminosity of $\sim$ 500 pb$^{-1}$ [45]
as well as by the detection efficiency and by the appropriate branching ratio $Br[\psi(2S) \to \mu^+\mu^-]\approx$ 0.77\%. Using a conservative detection efficiency of 10\%,
we estimate about of 115--2310 and 1155--34650 event numbers per year for the $\psi(2S)$ signal in the cases of the $^{12}$C and $^{184}$W target nuclei, respectively. We see that a sufficiently large number of $\psi(2S)$ events could be observed at well above threshold photon energies of 11.5--16.4 GeV. Hence, the $\psi(2S)$ excitation function measurements at these energies will allow for to set tight constraints on the $\sigma_{\psi(2S)N}$ cross section.

The momentum dependences of the absolute $\psi(2S)$ meson differential cross sections from direct productions processes
(1) and (2) in the ${\gamma}^{12}$C and ${\gamma}^{184}$W interactions, calculated on the basis of Eq. (19) for five
adopted values of the $\psi(2S)$--nucleon absorption cross section for laboratory angles of 0$^{\circ}$--10$^{\circ}$
and for initial photon energy of 13 GeV, are depicted in Fig. 4. It is clearly seen from this figure that the $\psi(2S)$
meson differential cross sections reveal a certain sensitivity
\footnote{$^)$Which is similar to that shown in Fig. 3.}$^)$
to this cross section mostly for the $^{184}$W target nucleus. Furthermore, in the latter case
the differential cross sections are roughly one order of magnitude larger than those on $^{12}$C nucleus and
they reach a rather measurable
\footnote{$^)$At upgraded up to 22 GeV CEBAF facility.}$^)$
strength $\sim$ 10--40 nb/(GeV/c) in the central momentum region of 11.5--12 GeV/c.
Thus, the $\psi(2S)$ meson differential cross section measurements at near-threshold photon energies $\sim$ 13 GeV
will open an additional possibility to constrain the $\psi(2S)N$ absorption cross section in cold nuclear matter.

The transparency ratios $S_A$ and $T_A$ of $\psi(2S)$ production from the direct reaction channels (1), (2)
in photon-induced reactions on nuclei
$^{12}$C, $^{27}$Al, $^{40}$Ca, $^{63}$Cu, $^{93}$Nb, $^{112}$Sn, $^{184}$W, $^{208}$Pb, and $^{238}$U
are plotted in Figs. 5 and 6, respectively, versus the mass number of the target nucleus
\footnote{$^)$It should be mentioned that the feasibility of the experimental determination of the width of a $D$ meson in a nuclear matter by using the method of the nuclear transparency $T_A$ has been studied in the very recent publication [74].}$^)$.
They have been calculated for the photon beam energy of 13 GeV in line with Eqs. (16) and (17), correspondingly,
and for five adopted values of the $\psi(2S)$--nucleon absorption cross section $\sigma_{\psi(2S)N}$.
It is clearly visible in Figs. 5 and 6 that the transparency ratios $S_A$ and $T_A$ show strong variations as
functions of this cross section and of the mass number. If we turn on absorption of $\psi(2S)$ mesons, then they drop
strongly along the target nuclei $^{12}$C -- $^{238}$U and, in particular, the transparency ratios $T_A$ reach values
of the order of 0.45 for heavy nuclei like $^{208}$Pb and $^{238}$U -- a large deviation from unity which should
be easily seen in a future experiment. In the case without $\psi(2S)$ absorption and with allowance for the nuclear effects we do not observe any significant decrease of the $T_A$ (and $S_A$) with increasing nuclear mass number $A$.
The sensitivity of the transparency ratios $S_A$ and $T_A$ to the input $\psi(2S)N$ absorption cross section is nicely
seen in Figs. 5 and 6.
Thus, there are a sizeable and measurable variations $\sim$ 16, 19, 23, 27\% in the ratio $S_A$
between calculations corresponding to the cases when the loss of flux of $\psi(2S)$ mesons in nuclei is determined by their absorption cross sections of 28 and 21 mb, 21 and 14 mb, 14 and 7 mb, 7 and 0 mb, respectively,  already for relatively "light" nuclei like $^{40}$Ca. For the medium-mass ($^{112}$Sn) and heavy ($^{238}$U) target nuclei these variations are even larger. They are about 21, 29, 31, 40\% and 23, 30, 44, 52\%, respectively. So, the highest sensitivity of the quantity $S_A$ to the cross section $\sigma_{\psi(2S)N}$ is observed for heavy target nuclei.
For the quantity $T_A$ the analogous variations are smaller but yet are experimentally distinguishable in the range of large A. They are about 4, 6, 7, 10\%, 8, 14, 15, 20\% and 11, 16, 26, 31\%, respectively, in the cases of "light", medium-mass and heavy target nuclei mentioned above.
Therefore, we can conclude that the observation of the A-dependences of the transparency ratios $S_A$ and $T_A$, at least, for large mass numbers $A$ in the future photoproduction experiments, for example, at the GlueX facility [49]
in JLab upgraded to 22 GeV would definitely allow to better constrain and to some degree discriminate among considered values of the $\psi(2S)N$ absorption cross section.

Another sources of information about this cross section are shown in Figs. 7 and 8 the photon energy and $\psi(2S)$ momentum dependences of the transparency ratio $T_A$ for the $^{184}$W/$^{12}$C combination calculated in line with Eqs. (17), (18) and (17), using the results presented in Figs. 2, 3 and 4, respectively.
One can see that the sensitivity of both dependences to the considered variations in the cross section $\sigma_{\psi(2S)N}$ is similar to that available in Fig. 6 for heavy nuclei. Hence, they can also be used for discriminating between possible options for the above cross section. Fig. 7 shows that for this aim we can employ,
in particular, the simple formula (18) at above threshold photon energies (at energies $E_{\gamma} > $ 10.93 GeV).

Using the simple formula (15) both for $\psi(2S)$ and for $J/\psi$ mesons at photon energies around of 11.5 GeV,
we also made predictions for the $A$-dependence of the $\psi(2S)$ transparency ratio $S_A^{\psi(2S)}$ normalized to
that $S_A^{J/\psi}$ for $J/\psi$ mesons
\footnote{$^)$It should be noted that the relative production of $\psi(2S)$ to $J/\psi$ states in high-energy proton-proton and PbPb collisions was a topic of study also for the recent LHCb experiments [73] and [35]. In addition, a few recent quarkonium results from the RHIC and LHC colliders in proton--proton, proton--nucleus and nucleus--nucleus collisions are reported very recently in Ref. [75].}$^)$
with the value of the $J/\psi$--nucleon absorption cross section $\sigma_{{J/\psi}N}=$ 3.5 mb motivated by the results from the $J/\psi$ photoproduction experiment at SLAC [39, 40]
for different values of the $\psi(2S)N$ absorption cross section $\sigma_{{\psi(2S)}N}$
\footnote{$^)$The use of the formula (15) for calculating $\psi(2S)$ transparency ratio $S_A^{\psi(2S)}$ is rather well
justified at photon energies around of 11.5 GeV as follows from the results presented in Figs. 2 and 3. Since these energies are far above the threshold energy for $J/\psi$ photoproduction on a free nucleon being at rest of 8.2 GeV, one can expect that this formula is also applicable at them for $J/\psi$ mesons.}$^)$.
They are shown in Fig. 9. The results, presented in this figure, have been obtained without accounting for in the calculations the $\psi(2S) \to J/\psi$ transition in nuclear medium and, nevertheless, could already be useful for getting first information about its strength here. Thus, we see that in the absence of such transition (or in the case of its quite negligible strength) the considered $\psi(2S)/J/\psi$ ratio, contrary to the results depicted in Fig. 6,
depends rather weakly on the nuclear mass number $A$ (or on the effective number of intranuclear nucleons involved in the interaction (cf. Ref. [36])), but the differences between all calculations corresponding to the adopted options for the cross section $\sigma_{\psi(2S)N}$, as before in this figure, are well separated and experimentally distinguishable.
Hence, a good agreement of the calculated $\psi(2S)/J/\psi$ ratio with the measured one will
enable us to get an additional information on this cross section and come to the conclusion
about reality of such scenario
\footnote{$^)$It should be noticed that in our calculations we did not taken into account the contribution to the $J/\psi$ production from the $\psi(2S)$s decays $\psi(2S) \to J/\psi+X$ outside of the nucleus, where $X$ could be two pions or other hadrons. Since the momenta of the $J/\psi$ mesons produced in these decays as well as in the direct
(${\gamma}+N \to {J/\psi}+N$) and in the two-step (${\gamma}+N \to {\psi(2S)}+N$, ${\psi(2S)}+N \to {J/\psi}+N$) processes are quite different, the decays $\psi(2S) \to J/\psi+X$ can be experimentally eliminated [43].}$^)$.
On the other hand, the presence of the strong $\psi(2S) \to J/\psi$ transition should result in the strong dropping with increasing $A$ behavior of the $\psi(2S)/J/\psi$ ratio, which can be considered as a signature of the strength of this transition. The calculations of this "new" ratio can be performed, using the results
presented both in this paper and in those [76, 77] where the corresponding model for the description of the two-step meson production processes in nuclear reactions has been developed. Since the experimental data on the above ratio are not available presently, the performing of such calculations is beyond the scope of the present paper.

In the ending of this paper we conclude that the information about the absolute total and differential cross sections for near-threshold photonuclear production of $\psi(2S)$ mesons as well as about their relative cross sections (transparency ratios) presented in this work would provide valuable assistance and guidance for future experimental studies aimed at exploring the $\psi(2S)$ absorption in cold nuclear matter.

\section*{4. Summary}

\hspace{1.5cm} In view of the expected data on the photonuclear production of the loosely bound $\psi(2S)$ mesons from the JLab upgraded to 22 GeV, in this work we have studied their inclusive photoproduction from nuclei near the kinematic threshold in the framework of the collision model, based on the nuclear spectral function, for incoherent direct photon--nucleon charmonium creation processes. The model takes into account the final $\psi(2S)$ absorption, target nucleon binding and Fermi motion. We have calculated the absolute and relative excitation functions for production
of $\psi(2S)$ mesons on $^{12}$C and $^{184}$W target nuclei at near-threshold photon beam energies of 8.0--16.4 GeV, the absolute momentum differential cross sections for their production off these target nuclei at laboratory polar angles of 0$^{\circ}$--10$^{\circ}$, the momentum dependence of the ratio of these cross sections as well as the A-dependences of the ratios (transparency ratios) of the total cross cross sections for $\psi(2S)$ production at photon energy of 13 GeV within the different scenarios for the $\psi(2S)N$ absorption cross section. We have also calculated the A-dependence of the ratio of $\psi(2S)$ and $J/\psi$ photoproduction transparency ratios at photon energies around of 11.5 GeV within the adopted scenarios for this cross section. We demonstrate that both the absolute and relative observables considered reveal a definite sensitivity to these scenarios. Therefore, the measurement of such observables in future experiments at the upgraded up to 22 GeV CEBAF facility near threshold might shed light both on the $\psi(2S)N$ absorption cross section and on its part associated with the nondiagonal transition $\psi(2S)+N \to J/\psi+N$ at finite momenta, providing
valuable insights for experimental studies of charmonium production and suppression in relativistic heavy-ion collisions.
\\

\end{document}